\documentclass[a4paper]{jpconf}
\usepackage{graphicx}
\usepackage[dvipdfm]{color}
\usepackage{taroMac}

\renewcommand{\Tkax}[1]{\ensuremath{\kappa _{\parallel #1}}}
\renewcommand{\TextRed}[1]{{#1}}
\renewcommand{\TextBlue}[1]{{#1}}
\renewcommand{\TextMagenta}[1]{{#1}}
\newif\ifHeader
\Headerfalse
\ifHeader
\usepackage{fancyhdr}

\newcommand{\TConfName}{
\begin{tabular}{r}
 27th International Conference on Low Temperature Physics (LT27) \\
 6-13, Aug., 2014, Palais Rouge, Buenos Aires, Argentina  \\
\end{tabular}
}

\makeatletter
\def\ps@myheadings{
  \let\ps@jpl@in\ps@plain
  \let\@oddfoot\@empty
  \let\@evenhead\@oddhead
  \def\@oddhead{\hfil\TConfName}
  \let\@mkboth\@gobbletwo
  \let\chaptermark\@gobble
  \let\sectionmark\@gobble}
\makeatother

\fi

\begin{document}

\title{Thermal Conductivity of the $S=1/2$ Quasi-One-Dimensional Ferromagnetic Spin System CsCuCl$_3$}

\author{
  T~Kawamata$^1$, 
  H~Sudo$^1$, 
  Y~Matsuoka$^1$, 
  K~Naruse$^1$, 
  M~Ohno$^1$, 
  T~Sasaki$^2$, 
  Y~Koike$^1$
}

\address{$^1$ Department of Applied Physics, Tohoku University, Sendai, \TextRed{980-8579} Japan}
\address{$^2$ Institute for Materials Research, Tohoku University, Sendai, \TextRed{980-8579} Japan}

\ead{tkawamata@teion.apph.tohoku.ac.jp}

\begin{abstract}
We have measured the thermal conductivity along the $c$-axis, \Tkax{c}, parallel to \TextRed{ferromagnetic} spin-chains of single crystals \TextRed{of the $S=1/2$ quasi-one-dimensional spin system CsCuCl$_3$} in magnetic fields up to 14 T, in order to investigate the thermal conductivity due to spins, \Tk{spin}, and the change of thermal conductivity corresponding to \TextRed{the change} of the spin state.  
In the temperature dependence of \Tkax{c}, no contribution of \Tk{spin} has been observed, while a dip has been observed at the antiferromagnetic phase transition temperature\TextRed{, \TT{N}}. 
Furthermore, it has been found that \Tkax{c} at \TextRed{a low temperature of} 5.1~K \TextRed{below \TT{N}} \TextRed{changes with increasing field perpendicular to the $c$-axis in good correspondence to the field-induced change} of the spin state. 
\end{abstract} 

%
%
\section{Introduction}
Thermal conductivity in low-dimensional quantum spin systems has attracted interest, 
\TextRed{owing to} the large \TextRed{contribution of the} thermal conductivity due to spins, \Tk{spin} \cite{Hess:EPJS151:2007:73,Sologubenko:JLTP147:2007:387}. 
Especially, a number of studies \TextRed{on} \Tk{spin} have been carried out in the \TextRed{spin quantum number} $S = 1/2$ one-dimensional (1D) antiferromagnetic (AFM) systems \cite{
Miike:JPSJ38:1975:1279,
Sologubenko:PRB64:2001:054412,
Sologubenko:EL62:2003:540,
Parfeneva:PSS46:2004:347,
Kawamata:JPSJ77:2008:034607,
Kawamata:JPCS200:2010:022023,
Hlubek:PRB81:2010:020405,
Kawamata:JPSJ83:2014:054601,
Matsuoka:JPSJ83:2014:064603}. 
The large contribution of \Tk{spin} has been observed as a peak or \TextRed{a} shoulder in \TextRed{the} temperature dependence of \TextRed{the} thermal conductivity along spin chains together with the contribution of \TextRed{the} thermal conductivity due to phonons, \Tk{phonon}.  
In the 1D ferromagnetic \TextRed{(FM)} systems, on the other hand, \TextRed{the} contribution of \Tk{spin} has not been observed clearly, 
\TextRed{though several studies on \Tk{spin} have been performed}
\cite{Gronckel:PRB37:1988:9915,Gronckel:PRB44:1991:4654,Kudo:JMMM272:2004:94,Choi:JMMM272:2004:970,Cheng:PRB79:2009:184414}. 

\TextRed{The compound} \TABX{Cs}{Cu}{Cl} \TextRed{belongs to} the \TextRed{so-called} $ABX_3$ \TextRed{family including} well-known quasi-1D \TextRed{spin} systems with frustration. 
The compound contains linear spin-chains of face-sharing CuCl$_6$ octahedra running along the $c$-axis. 
The spin chains separated by Cs$^{+}$ ions form an equilateral triangular lattice in the $c$-plane. 
The intrachain exchange interaction $J$ is \TextRed{FM} and \TextRed{estimated as $\sim -28~\TK$, while} the interchain exchange interaction $J'$ \TextRed{is AFM and} estimated \TextRed{as} $\sim 5~\TK$ \cite{Mekata:JMMM140:1995:1987}. 
\TextRed{Therefore, frustration is expected to exist more or less between the spin chains. }
The CuCl$_6$ octahedra \TextRed{are} distort\TextRed{ed due to} \TextMagenta{the} Jahn-Teller effect, \TextRed{so that} the Dzyaloshinsky-Moriya interaction \TextRed{is induced} between spins along the $c$-axis. 
\TextRed{An AF long-range order appears at the phase transition temperature, \TT{N}, $\sim 10.7~\TK$, owing to $J'$} \cite{Adachi:JPSJ49:1980:545}. 
\TextRed{The spin structure at low temperatures below \TT{N} is} helical with \TextRed{the} pitch of $\sim 5.1\Tdeg$ along the $c$-axis \TextRed{owing to the} combination of $J$ and the Dzyaloshinsky-Moriya interaction. 
In the $c$-plane, the \TextRed{so-called} $120\Tdeg$ spin structure is formed \TextRed{owing to} $J'$ and 
the spin \TextRed{directions are confined} in the $c$-plane \cite{Adachi:JPSJ49:1980:545}.
\TextRed{Moreover, it has been known that the pitch decreases by the application of magnetic field perpendicular to spin chains \cite{Mino:PB201:1994:213,Nojiri:PB241:1997:210,Stuesser:JPCM14:2002:5161}}. 

\TextRed{In this paper,} we have measured the thermal conductivity along the $c$-axis, \Tkax{c}, parallel to spin chains of \TABX{Cs}{Cu}{Cl} \TextRed{in order to investigate \TextMagenta{the} contribution of \Tk{spin}}. 
Furthermore, 
\TextRed{we have also investigated the relation between the spin state and the behavior of the thermal conductivity, because} 
thermal conductivity is a good probe detecting a change of the spin state in a spin system \TextRed{via the change of} 
the scattering rate of heat carries \cite{
Kudo:JPSJ70:2001:1448,
Kudo:JMMM272:2004:214,
Sato:JPCS200:2010:022054,
Li:PRB87:2013:214408,
Zhang:PRB89:2014:094403}.

%
%
\section{Experimental}
Single crystals of \TABX{Cs}{Cu}{Cl} were grown from solution. 
Thermal conductivity measurements were carried out by the conventional steady-state method in magnetic fields up to 14~T. 

%
%
\section{Results and Discussion}
\begin{figure}[t]
	\includegraphics[width=18pc]{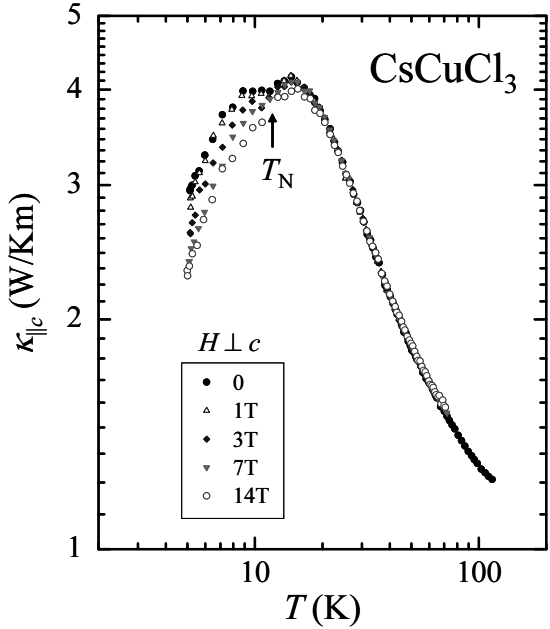}
	\hspace{2pc}
	\begin{minipage}[b]{14pc}
		\vspace{-3pc}
		\caption{Temperature dependence of the thermal conductivity along the $c$-axis, \Tkax{c}, parallel to spin chains in zero field and magnetic fields perpendicular to the $c$-axis. }
		\label{fig:kappa}
	\end{minipage}
\end{figure}

Figure \ref{fig:kappa} shows the temperature dependence of \Tkax{c} parallel to spin chains in \TextRed{zero field and} magnetic fields perpendicular to the $c$-axis \TextRed{at low temperatures} below $\sim 100$~K. 
It \TextBlue{has been} found that \Tkax{c} show a peak around $10$~K with a dip at \TT{N} in zero field. 
\TextRed{This kind of dip at} \TT{N} has been observed in several antiferromagnets \cite{Sologubenko:EL62:2003:540,Slack:PRL1:1958:359,Lewis:JPCSSP6:1973:2525}, 
\TextRed{and is interpreted as being} 
due to \TextRed{the} strong scattering of heat carries caused by the critical fluctuations \TextRed{around} \TT{N} or due to the increase of the mean free path of heat carries just below \TT{N} 
\TextRed{on account of} 
the marked reduction of 
\TextRed{the} 
scattering caused by the development of the long-rage order. 
Therefore, this behavior of \Tkax{c} in zero field is likely to be \TextRed{mainly} due to \TextRed{the} contribution of \Tk{phonon} 
\TextRed{affected by the scattering of phonons} 
by magnetic excitations. 
The contribution of \Tk{spin} cannot be observed clearly in \TABX{Cs}{Cu}{Cl}. 
Recently, 
single crystals \TextRed{of \TABX{Cs}{Cu}{Cl} \TextMagenta{of} a single domain with \TextMagenta{homochirality} of spins have been grown by Kousaka \textit{et al}. \cite{Kousaka:JPCS502:2014:012019}.} 
\TextRed{On the other hand,} 
our single crystals \TextRed{of the} zigzag shape \TextRed{may be of multi-domains}. 
Therefore, \Tk{spin} \TextRed{might} strongly \TextRed{be suppressed due to the strong scattering of magnons at the domain boundaries}. 

\begin{figure}[t]
	\hspace{1pc}
	\includegraphics[width=18pc]{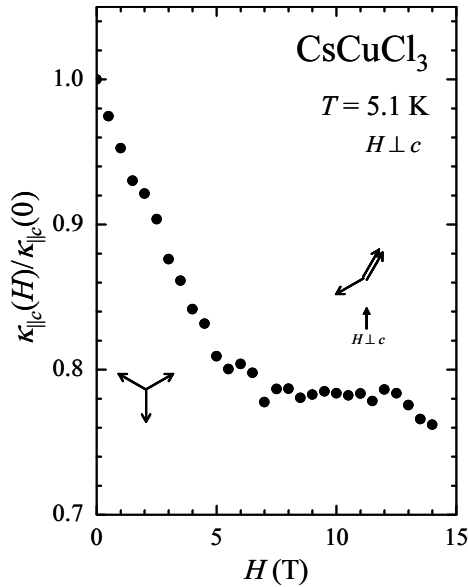}
	\hspace{-1pc}
	\begin{minipage}[t]{15pc}
		\vspace{-11pc}
		\caption{Magnetic-field dependence of 
		\TextRed{the thermal conductivity along the $c$-axis normalized by the value in zero field,} 
		$\Tkax{c}(H)/\Tkax{c}(0)$, of CsCuCl$_3$ in magnetic fields perpendicular to the $c$-axis. 
		\TextBlue{Three arrows on the left and right sides mean $120\Tdeg$ and coplanar 2-1-type spin structures, respectively. }
		}
		\label{fig:kH}
	\end{minipage}
\end{figure}

By the application of magnetic field perpendicular to the $c$-axis, 
\TextRed{it has been found that} \Tkax{c} \TextRed{at low temperatures} below \TT{N} is suppressed, as shown Fig. \ref{fig:kappa}. 
To see the suppression in detail, 
the magnetic-field dependence of \TextRed{\Tkax{c} normalized by the value in zero field,} 
$\Tkax{c}(H)/\Tkax{c}(0)$, at $5.1~\TK$ \TextRed{is shown in Fig. \ref{fig:kH}}. 
It \TextBlue{has been} found that $\Tkax{c}(H)/\Tkax{c}(0)$ \TextRed{monotonically} decreases \TextRed{with increasing field} up to $\sim 9$~T, 
becomes constant between $\sim 9$~T and $\sim 12$~T, 
and then decreases again \TextRed{above $\sim 12$~T.} 
This behavior is very similar to the magnetic-field dependence of \TextRed{the} incommesurabiltiy, $\delta$, which is proportional to the pitch of the helical spin structure along the $c$-axis 
\cite{Mino:PB201:1994:213,Nojiri:PB241:1997:210,Stuesser:JPCM14:2002:5161}. 
\TextRed{It is known that} 
the helical and $120\Tdeg$ spin structures are 
\TextRed{maintained in low magnetic fields below $\sim 9$~T, though the pitch of the helical structure along the $c$-axis decreases with increasing field} \cite{Stuesser:JPCM14:2002:5161}. 
\TextRed{The $120\Tdeg$ spin structure in the $c$-axis is expected to become unstable by the application of magnetic field in the $c$-plane.} 
\TextRed{Therefore, }
it is likely that \Tkax{c} is 
\TextRed{suppressed with increasing field} 
due to 
\TextRed{the enhancement} 
of \TextRed{the} magnetic \TextMagenta{fluctuations} scatter\TextRed{ing} phonons. 

In a region between 9~T and 16~T, 
\TextRed{it is known that} 
the helical and 
\TextRed{the so-called} 
coplanar 2-1-type spin structures, 
where two \TextRed{thirds of} spins are \TextRed{roughly} parallel and the \TextRed{others are roughly anti}parallel to the magnetic-field \TextRed{direction} \TextMagenta{as shown in Fig.2}, 
are formed. 
The constant value of $\Tkax{c}(H)/\Tkax{c}(0)$ in this region implies that the number of magnetic excitations scattering phonons does not change \TextMagenta{so much}. 
This is consistent with \TextRed{the} locking \TextRed{phenomenon} of the \TextRed{spin} structure 
\TextRed{in this region observed in} the neutron scattering \TextRed{experiment} \cite{Stuesser:JPCM14:2002:5161} \TextRed{and with the} plateau-like behavior \TextRed{in this region} observed in the magnetization curve \TextRed{also} \cite{Nojiri:JPC49:1988:C8-1459}. 

\TextRed{In high magnetic fields} above $\sim 12$~T, on the other hand, the suppression of \Tkax{c} can be explained by the \TextRed{enhancement} of magnetic fluctuations \TextRed{scattering phonons, because the} 
field-induced transition to \TextRed{the non-helical} commensurate phase 
\TextRed{with $\delta = 0$ takes place at $\sim 16$~T \cite{Nojiri:PB241:1997:210,Stuesser:JPCM14:2002:5161}. } 
\TextRed{The behavior of \Tkax{c}} is also \TextRed{roughly} consistent with the neutron scattering result that \TextRed{the} value of \TextRed{$\delta$ starts to} decrease \TextRed{at} $\sim 14$~T \TextRed{with increasing field} \cite{Nojiri:PB241:1997:210,Stuesser:JPCM14:2002:5161}. 

%
%
\section{Conclusions}
We \TextRed{have} measured \TextRed{\Tkax{c}} parallel to spin chains of CsCuCl$_3$, in order to investigate \TextRed{the presence or absence of} \Tk{spin} and \TextRed{the change of the} spin state. 
Neither peak nor shoulder \TextRed{due to the contribution of} \Tk{spin} \TextRed{has been} observed in \TextRed{the temperature dependence of} \Tkax{c} \TextRed{in zero field}. 
This may be due to the strong scattering \TextRed{of magnons at} boundaries 
\TextRed{of} domains \TextRed{with homochirality of spins in the}
 helical spin structure \TextRed{along the $c$-axis}. 
\TextRed{By the application of magnetic field perpendicular to the $c$-axis up to 14~T, it has been found that} \Tkax{c} 
\TextRed{at 5.1~K below \TT{N} decreases with increasing field up to $\sim 9$~T, is constant between $\sim 9$~T and $\sim 12$~T, and} decreases again above $\sim 12$~T. 
These behaviors 
\TextRed{have been explained as being due to the change of \Tk{phonon} caused by the change of the scattering of phonons by magnetic excitations due to}
the \TextRed{filed-induced} change of the \TextRed{spin state}. 

%
%
\vspace{-0.5pc}
\ack
The thermal conductivity measurements were performed at the High Field Laboratory for Superconducting Materials, Institute for Materials Research, Tohoku University. 
This work was supported by a Grant-in-Aid for Scientific Research from the Ministry of Education, Culture, Sports, Science and Technology, Japan.

%
%
\vspace{2pc}
\providecommand{\newblock}{}

\end{document}